\documentclass[aps,prl,twocolumn,superscriptaddress,preprintnumbers,nofootinbib]{revtex4-1}
\usepackage{amsmath,amssymb,graphicx,natbib,bm}
\usepackage{enumerate}
\usepackage{hyperref}
\usepackage{breakurl}
\usepackage{color}

\newcommand{\acro}[1]{\textsc{\MakeLowercase{#1}}} 
\newcommand{\osn}{\oldstylenums}

\def\be{\begin{equation}}
\def\ee{\end{equation}}
\def\ba{\begin{eqnarray}}
\def\ea{\end{eqnarray}}

\usepackage[table]{xcolor}
\usepackage[utf8]{inputenc} 

\begin{document} 

\title{Hydrogen portal to exotic radioactivity}

\author{David McKeen}
\email{mckeen@triumf.ca}
\affiliation{TRIUMF, 4004 Wesbrook Mall, Vancouver, BC V6T 2A3, Canada}

\author{Maxim Pospelov}
\email{pospelov@umn.edu}
\affiliation{School of Physics and Astronomy, University of Minnesota, Minneapolis, MN 55455, USA}
\affiliation{William I. Fine Theoretical Physics Institute, School of Physics and Astronomy, University of Minnesota, Minneapolis, MN 55455, USA}

\author{Nirmal Raj}
\email{nraj@triumf.ca}
\affiliation{TRIUMF, 4004 Wesbrook Mall, Vancouver, BC V6T 2A3, Canada}

\begin{abstract}
We show that in a special class of dark sector models, the hydrogen atom can serve as a portal to new physics, through its decay occurring in abundant populations in the Sun and on Earth.
The large fluxes of hydrogen decay daughter states can be detected via their decay or scattering.
By constructing two models for either detection channel, we show that the recently reported excess in electron recoils at \acro{XENON}\osn{1}\acro{T} could be explained by such signals in large regions of parameter space unconstrained by proton and hydrogen decay limits.
\end{abstract}

\date{\today}

\maketitle


{\bf \em Introduction.}---The extremely successful experimental program of building ever larger and cleaner 
dark matter search experiments based on two-phase xenon has led to 
extremely important advances in particle astrophysics. Many dark matter models, 
including both heavy and light particles, are considerably constrained or excluded. 
At the same time, these experiments -- by their sheer size and low levels of contamination -- prove to be universal tools in detecting {\em any} 1-100\,keV
energy deposition. 
Reaching counting rates as low as $O(10-100){\rm /ton/yr/keV}$,
the \acro{XENON}\osn{1}\acro{T} experiment has shown itself to be a leader in probing 
rare processes in the energy domain inaccessible to 
conventional neutrino experiments. 

In addition to setting important constraints, 
the \acro{XENON}\osn{1}\acro{T} experiment has observed an excess over known backgrounds in electron recoils in 2--3 keV energies in 0.65 ton-year of data~\cite{X1TExcess}.
Putative explanations proposed 
include an unmeasured tritium  background~\cite{X1TExcess,robinson2020xenon1t} (or any other missed radioactive background in the keV 
range),
and physics scenarios beyond the Standard Model (\acro{sm})~\cite{
X1TExcess,
Kannike:2020agf,
Takahashi:2020bpq,
*DiLuzio:2020jjp,
*Alonso-Alvarez:2020cdv,
*Boehm:2020ltd,
*Bally:2020yid,
*AristizabalSierra:2020edu,
*1802520,
*Fornal:2020npv,
*Du:2020ybt,
*Chen:2020gcl,
*1802522,
*1802529,
*Harigaya:2020ckz,
*Su:2020zny,
*Bell:2020bes,
*Buch:2020mrg,
*1802563,
*Choi:2020udy,
*Paz:2020pbc,
*1802561,
*an2020new,
*baryakhtar2020electromagnetic,
*bramante2020electric,
*gao2020reexamining,
*lindner2020xenon1t,
*zu2020mirror,
*bloch2020exploring}.
These papers explore a diverse set of ideas on how the unforeseen keV-scale events 
may emerge in various new physics scenarios. 
The following broad picture has emerged with respect to to the explanation of the excess thus far: 
\begin{itemize}
    \item At the moment, there is no clear model of light sub-keV particles that would be emitted from the Sun, and absorbed by the detector without running into other problems such as astrophysical energy loss constraints.
    
    \item Light neutrinos, produced in the Sun, Earth or reactors, cannot be the source of the signal without deviating from the conventional \acro{SM} physics. 
    Even then many scenarios involving novel interactions such as {\em e.g.}  light neutrino magnetic moments are again ruled out by stellar energy loss constraints and/or cosmology. 
    \item So far the dark matter route seems to be theoretically the easiest avenue 
    for arranging excess events. Ideas include absorption of keV scale dark matter, inelastic dark matter de-exciting in the detector, or elastic scattering of a dark matter sub-component that moves faster than expected virial Galactic velocities. In many cases the astrophysics constraints can be relaxed or avoided
    because the dark matter particles are heavy enough, while the absorption of light keV dark matter can also be safe in many scenarios due to extremely large number densities, and correspondingly tiny couplings of such particles.

\end{itemize}

In this paper we explore the idea of radioactivity induced by standard hydrogen in models where the electron and proton couple to exotic particles. 
This may be viewed as an alternative to using dark matter as a source of keV energy. 
Due to the GeV-scale mass of daughter particles many of the astrophysics constraints are irrelevant, while hydrogen is guaranteed to be an abundant source, both on Earth and in the Sun, even after taking into account strong constraints on its lifetime.

It is known that dark matter decays may lead to a substantial
flux of particles at Earth's location, with the main sources 
of the flux being the local Milky Way halo as well as the global dark matter density in the Universe \cite{Cui:2017ytb}. For a maximum allowed abundance of dark matter particles, and shortest lifetime scales 
(exceeding the lifetime of the universe $\tau_{\rm U}$ by a factor of $\sim 10 $), the flux of daughter particles can achieve substantial values and even be comparable with the solar or Earth neutrino fluxes, or even those of the cosmic neutrino background~\cite{McKeen:2018xyz,*Chacko:2018uke}.
Moreover, even the weak-strength interactions of daughter products could be detected using the most sensitive neutrino and dark matter detectors, existent or in development~\cite{Cui:2017ytb,McKeen:2018xyz,*Chacko:2018uke}. 
While the decaying dark matter scenario is full of ``unknown unknowns,'' their number shrinks if we get to possible decays of \acro{SM} particles. 
The most abundant long-lived particle with appreciable mass that may decay without violating charge conservation is the proton.
The maximum flux of daughter particles for $O(1)$ multiplicity can be estimated as $N_{\rm p}^{\odot}
\tau_{p}^{-1} (4\pi {\rm A.U.}^2)^{-1}$, where $\tau_{p}$ is the experimental limit on the proton lifetime and $N_{\rm p}^{\odot}$ is the total number of protons in the Sun. 
An interesting caveat that has been discussed in recent literature is that if the masses of new physics daughter particles lie close to $m_{p,n}$, the proton itself may be stable but the {\em hydrogen atom}
could decay. 
Although relatively strong bounds can be derived on the lifetime of H using the sensitivity to the final state photon in the {\em radiative} capture of $e^-$ by $p$~\cite{McKeen:2020zni}, these bounds are in fact many orders of magnitude milder than $\tau_{p}$ limits. 
To summarize, one has the following picture of the maximum attainable fluxes of daughter particles 
(assuming $O(1)$ multiplicities):
\begin{equation}
\begin{aligned}
\Phi^{\rm global+MW}_{\rm DM~decay} &\sim 10^{4}~{\rm cm^{-2}s^{-1}}
\bigg( \frac{10\,\tau_{\rm U}}{\tau_{\rm DM}}\bigg)  \bigg(\frac{1\,{\rm GeV}}{m_{\rm DM}}\bigg)~, \\
\Phi^\odot_{\rm proton~decay} &\sim  10^{-8}~{\rm cm^{-2}s^{-1}} \bigg(\frac{10^{30}\,{\rm yr}}{\tau_{p}}\bigg)~,\\
\Phi^\odot_{\rm H~decay}  &\sim
10^{3}~{\rm cm^{-2}s^{-1}} \bigg(\frac{10^{28}\,{\rm s}}{\tau_{\rm H}}\bigg)~,\\
\Phi^\oplus_{\rm H~decay}  &\sim
1~{\rm cm^{-2}s^{-1}} \bigg(\frac{10^{28}\,{\rm s}}{\tau_{\rm H}}\bigg).
\label{fluxes}
\end{aligned}
\end{equation}

\begin{figure}
    \centering
    \includegraphics[width=0.46\textwidth]{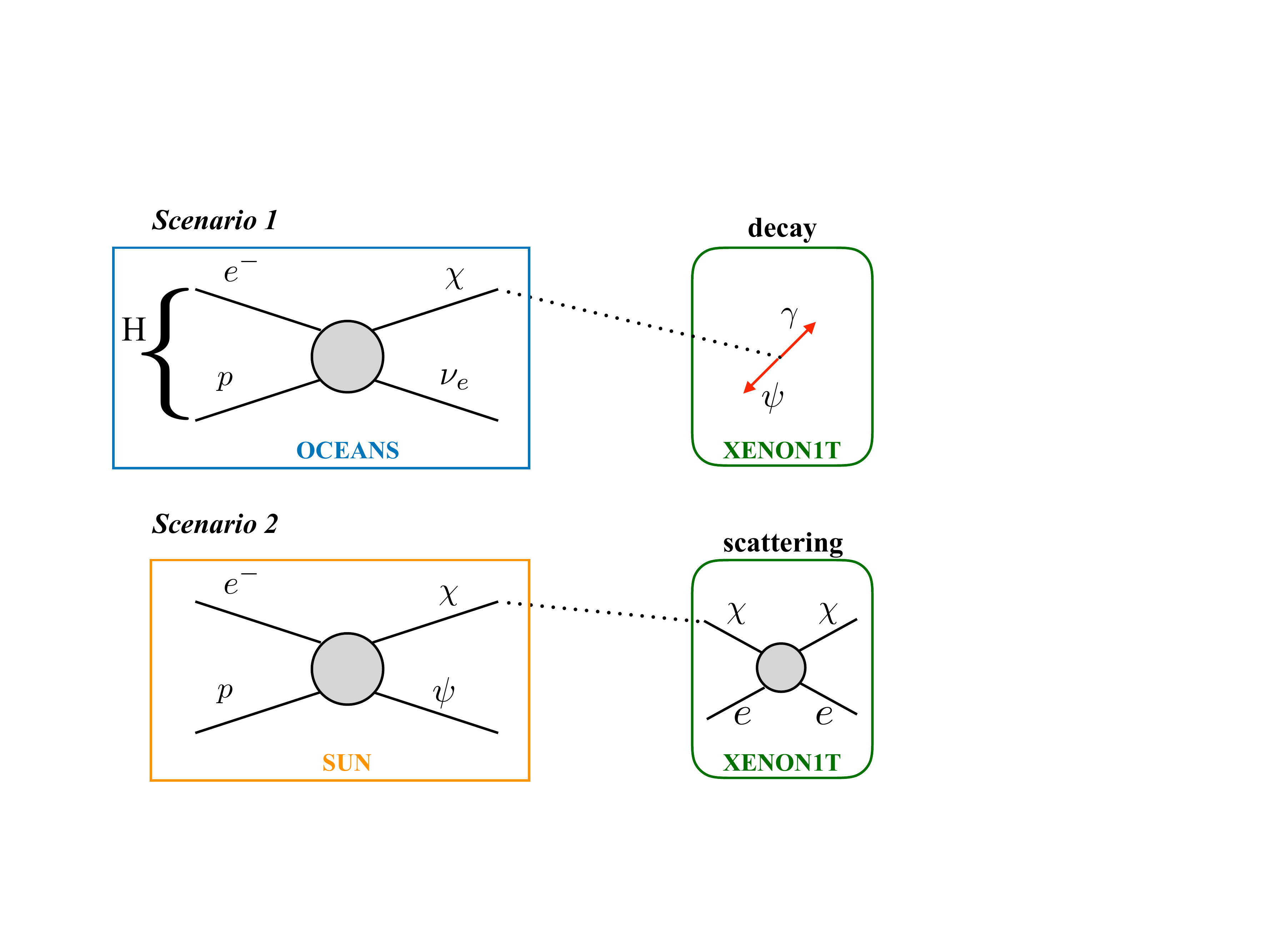} 
    \caption{The two scenarios considered in this work that could explain the \acro{XENON}\osn{1}\acro{T} electron recoil excess.
    In the first, H atoms decay on Earth (primarily in the oceans) to produce a dark baryon that propagates through the Earth and decays in a detector to another fermion and a photon.
    In the second, $e^-p$ capture in the Sun produces a fast-moving long-lived dark sector particle that scatters on detector electrons.
     }
    \label{fig:scenarios}
\end{figure}

All of these decays, and we will concentrate on that of H, 
can be considered as {\em portals} to a new sector 
that may have additional interactions with the \acro{SM}, and reveal themselves via scattering or decay. Motivated by recent 
\acro{XENON}\osn{1}\acro{T} results, we will focus on 
sub-10 keV energy deposition, while in principle H-portal couplings could lead to up to few 100 MeV energy release. 
We investigate the detection of large fluxes of daughter states of H decay occurring in both the Sun and Earth.  
In particular we explore the two scenarios summarized in Fig.~\ref{fig:scenarios}: the daughter states could either scatter on electrons or decay to a photon final state in the detector.
We show that in both scenarios there exist large regions of parameter space that result in the right flux and energy deposition in \acro{xenon}\osn{1}\acro{t} to fit the recent excess well. From a more conservative standpoint, independent of this particular excess, sensitive  neutrino and direct dark matter detection experiments can be used to probe properties of particles emerging from H decays.

\begin{figure}
    \centering
    \includegraphics[width=0.54\textwidth]{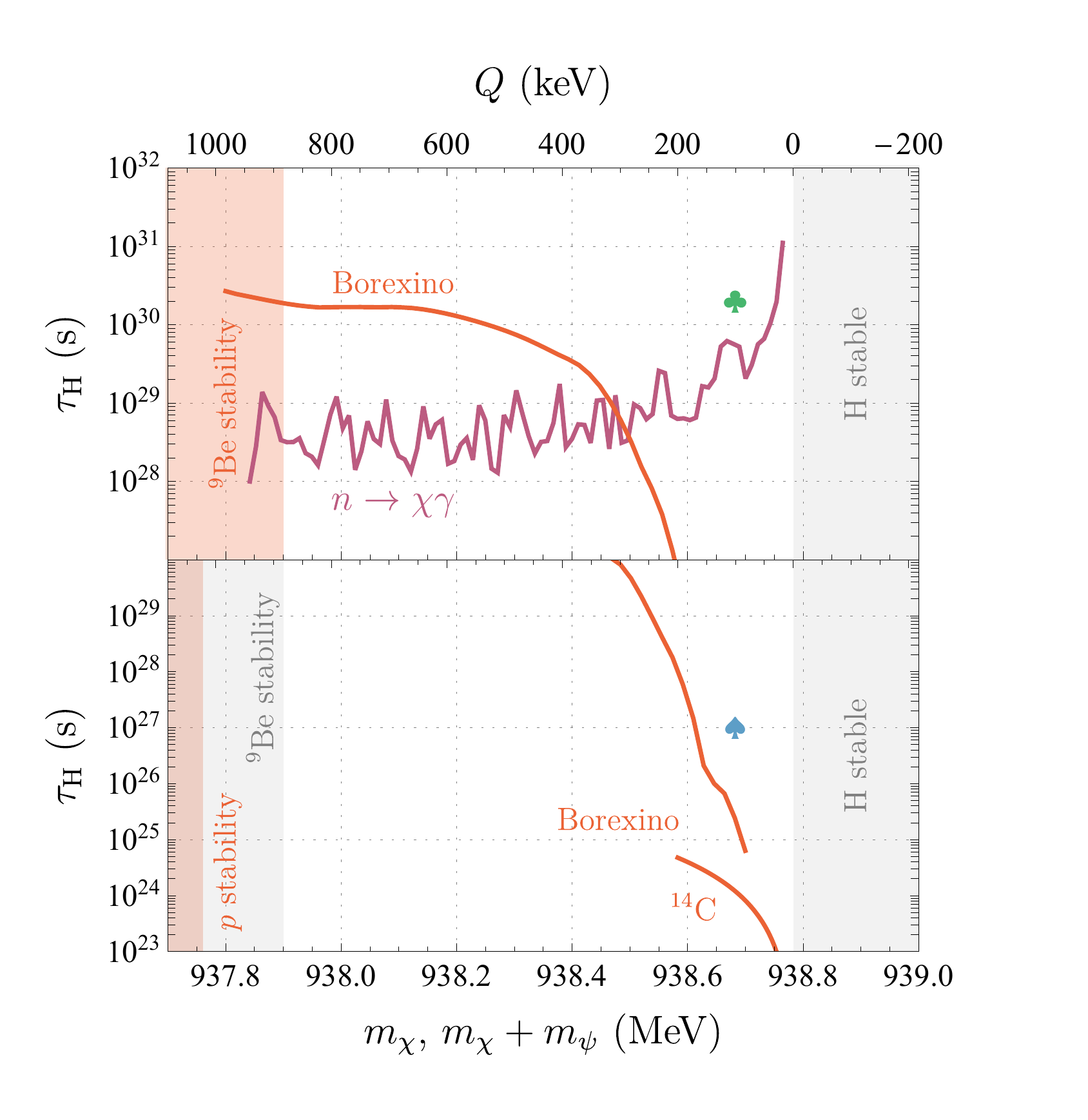} 
    \caption{
    Lower limits, as derived in \cite{McKeen:2020zni}, on the H lifetime as functions of $Q$ and the total mass of the decay final state in our two scenarios.
{\bf Top}: Limits from the low energy Borexino spectrum~\cite{Agostini:2015oze} (red) and the search for $n\to\chi\gamma$~\cite{Tang:2018eln} (purple) in the neutron-mixing model of Scenario 1.
{\bf Bottom:} Limits from \cite{Agostini:2015oze} and a study of $^{14}{\rm C}$ purity~\cite{Alimonti:1998rc} on the \acro{EFT} interaction of Scenario 2. 
 Also shown with a green club and blue spade are the benchmark ($\tau_{\rm H}$, $Q$) in Scenario 1 and 2, respectively, that give rise to the spectra in Fig.~\ref{fig:signals}.}
    \label{fig:tauH}
\end{figure}

{\bf \em  Terestrial $\rm H$ decay to metastable state.}---We first consider a model with the neutron portal, in which the neutron mixes with a dark neutron $\chi$ (that may be  elementary or composite):
\begin{equation}
    {\cal L}\supset \delta\left(\bar n\chi+\bar\chi n\right)
\end{equation}
with mixing angle $\theta=\delta/\left(m_n-m_\chi\right)$. 
This model has received much theoretical and experimental 
attention in connection with the 
neutron lifetime discrepancy~\cite{Fornal:2018eol,Berezhiani:2018udo} as well as potential connections to dark matter~\cite{McKeen:2015cuz,Allahverdi:2017edd,*Karananas:2018goc}. 
If $m_\chi<m_{\rm H}\simeq m_p+m_e$, atomic hydrogen decays to $\chi+\nu_e$ with a lifetime~\cite{Berezhiani:2018udo,McKeen:2020zni}
\begin{equation}
    \tau_{\rm H}=10^{29}~{\rm s}\left(\frac{10^{-10}}{\theta}\right)^2\left(\frac{m_e}{Q}\right)^2~,
\end{equation}
where $Q=m_{\rm H}-m_\chi$. 
In this model the hydrogen lifetime through this channel is constrained to be $\tau_{\rm H}\gtrsim 10^{29}~\rm s$ by a search for $n\to\chi\gamma$~\cite{Tang:2018eln} and a recast of Borexino data~\cite{Agostini:2015oze} to constrain $H\to\chi\nu\gamma$~\cite{McKeen:2020zni}, displayed in the top panel of Fig.~\ref{fig:tauH}.
There is also a lower limit on the $\chi$ mass of $m_\chi>938.0~\rm MeV$ from the stability of $^9{\rm Be}$~\cite{McKeen:2015cuz,Pfutzner:2018ieu}. 
Given this bound, the speed of $\chi$'s produced in H decay is $v_\chi/c\simeq Q/m_\chi<0.8\times 10^{-3}$. 
This is lower than the Sun's surface escape velocity of $0.002c$, which can be overcome only by the energetic tail of the Boltzmann distribution. 
Therefore for detecting $\chi$ on Earth, we must consider terrestrial sources of hydrogen. 
One simple possibility to detect $\chi$'s after production involves a slight extension of the model: we add a neutral fermion $\psi$, and a transition magnetic dipole moment between $\chi$ and $\psi$,\footnote{Indeed we expect that the dark neutron has additional interactions because of constraints from neutron stars~\cite{McKeen:2018xwc,*Baym:2018ljz,*Motta:2018rxp,*
Cline:2018ami} as well as inheriting a magnetic moment from the neutron.}
\begin{equation}
    {\cal L}\supset \frac{1}{\Lambda_d}\bar\psi \sigma^{\mu\nu} F_{\mu\nu} \chi~.
\end{equation}
If $m_\chi-m_{\psi}\equiv\delta m>0$, $\chi$ decays to $\psi$ and a photon of energy $\omega=\delta m$  with the decay length 
\begin{equation}
\begin{aligned}
    v_\chi\tau_\chi&=6\times 10^{8}~{\rm cm}\left(\frac{3~\rm keV}{\delta m}\right)^3
    \\
    &\quad\quad\times\left(\frac{\Lambda_d}{10^5~\rm GeV}\right)^2\left(\frac{Q}{100~\rm keV}\right)
\end{aligned}
\end{equation}
when produced in H decay.
For $v_\chi\tau_\chi$ not too small compared to the radius of Earth, $R_\oplus=6.4\times 10^8~\rm cm$, the decay of terrestrial atomic hydrogen can produce photons through $\chi$ decay that can potentially be observable.

\begin{figure}
    \centering
    \includegraphics[width=0.46\textwidth]{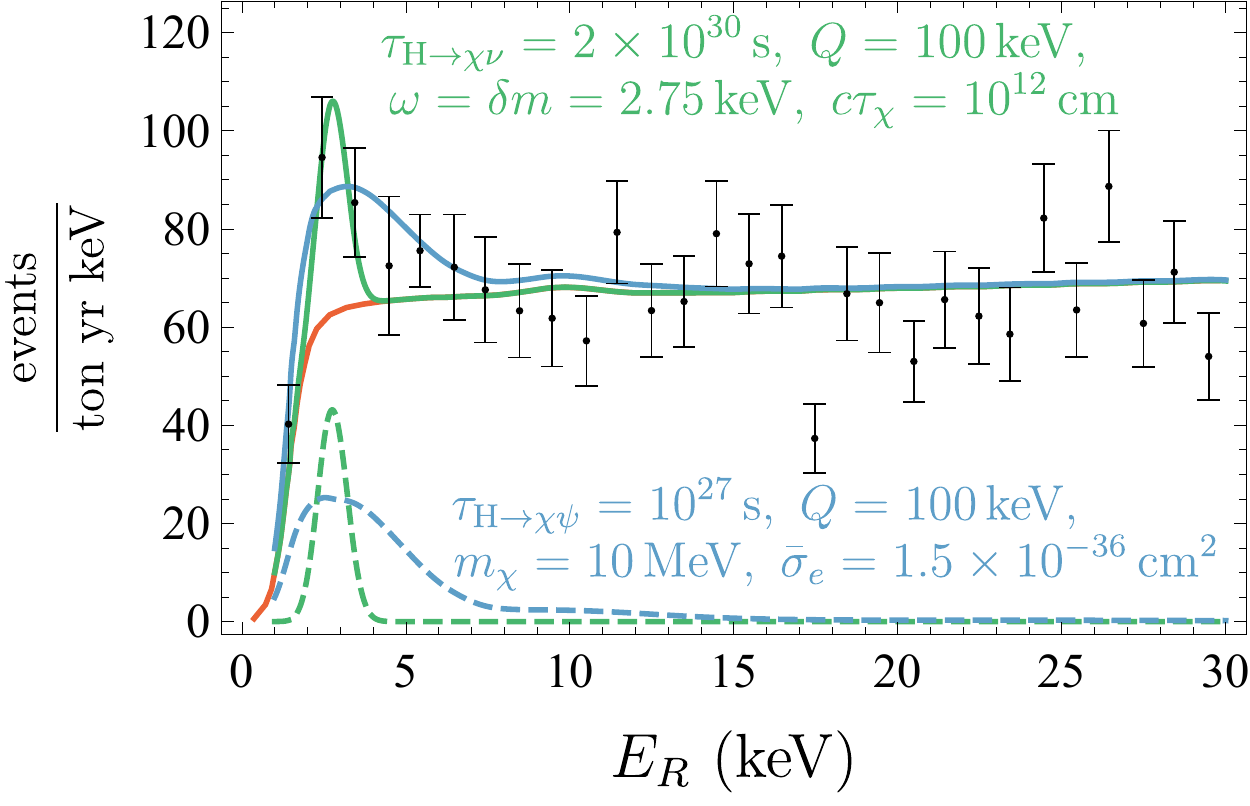} 
    \caption{
Event rates in \acro{xenon}\osn{1}\acro{t} in the scenarios of decay (green) and scattering (blue) of hydrogen daughters, seen to significantly reduce tensions in the data for the benchmark points indicated in Fig.~\ref{fig:tauH}.
Here the red curve denotes the known backgrounds taken from \cite{X1TExcess}. The dashed curves depict the signal only while the solid curves depict the signal plus background.
See text for further details.}
    \label{fig:signals}
\end{figure}

The vast majority of atomic hydrogen on Earth is in the ``hydrosphere''---the Earth's collection of water---with a total volume of $1.4\times 10^9~\rm km^3$~\cite{hydrosphere}, 
corresponding to $N_{\rm H}\simeq 10^{47}$ hydrogen atoms. 
Modeling the hydrosphere as uniformly covering the Earth's surface, the rate of $\chi$ decays in a detector of volume $\ell_{\rm det}^3$ near the surface is
\begin{equation}
    \begin{aligned}
    R&=\frac{f_{\rm mol} N_{\rm H}}{\tau_{\rm H}}\frac{\ell_{\rm det}^3}{4\pi R_\oplus^3}F\left(\frac{R_\oplus}{\ell_\chi},\frac{r_{\rm min}}{R_\oplus}\right)~,
    \end{aligned}
\end{equation}
where $\ell_\chi=(1-v_\chi^2)^{-1/2} v_\chi\tau_\chi\simeq v_\chi\tau_\chi$ is the mean distance travelled by $\chi$'s produced in H decay, $r_{\rm min}$ is the distance from the detector to the nearest point in the hydrosphere, $f_{\rm mol}$ characterizes the difference between the atomic and molecular H lifetimes arising from the distortion of the molecular electron wavefunction, and
\begin{equation}
    \begin{aligned}
    F(y,\delta)=\frac{y}{2}\int_{-1}^{1-\delta^2/2} dx\frac{\exp\left[-y\sqrt{2(1-x)}\right]}{2(1-x)}.
    \end{aligned}
\end{equation}
For $\ell_\chi\sim R_\oplus,\,r_{\rm min}$, $F$ is ${\cal O}(1)$ and an appreciable number of $\chi$'s produced terrestrially can decay in an experiment on Earth. 
In a liquid Xe detector such as \acro{XENON}\osn{1}\acro{T}, this gives a photon deposition rate of
\begin{align}
    R&\simeq\frac{140}{\rm ton~ yr}\left(\frac{10^{30}~{\rm s}}{\tau_{\rm H}}\right)\left(\frac{f_{\rm mol}}{0.5}\right)\times F~.
\end{align}
Photons with $\omega\sim{\rm keV}$ produced in \acro{XENON}\osn{1}\acro{T} have a short travel time 
before they initiate photo-absorption that eventually leads to multiple ionization electrons~\cite{Aprile:2019dme}.
This estimate shows that in the neutron-mixing model with a value of $\tau_{\rm H}$ allowed by data and $v_\chi\tau_\chi\sim R_\oplus$ (such that $F$ is not small),  hydrogen decay on Earth remarkably produces the right amount of photons at the right energy to explain the \acro{XENON}\osn{1}\acro{T} excess in Ref.~\cite{X1TExcess}. 
In Fig.~\ref{fig:signals} we show the total event rate for $\omega=\delta m = 2.75~{\rm keV}$ photons, with $\tau_{\rm H} = 2 \times 10^{30}$~s, $Q = 100$~keV, $\ell_\chi=3\times 10^{12}~{\rm cm}$ (corresponding to $\Lambda_d=1.1\times10^5~{\rm GeV}$), $f_{\rm mol}=0.5$, and $r_{\rm min}/R_\oplus=0.2$ which produces about 50 events after convolving with the detection efficiencies given in Ref.~\cite{X1TExcess}.
Here we account for the detector energy resolution of 0.45~keV~\cite{X1Tres} by Gaussian-smearing the signal rate.
The photon deposition rate for our benchmark point is seen to fit the excess well, borne out by a 2-parameter fit giving $\chi^2$/d.o.f. = 0.75 with a $\Delta \chi^2$ of 10.77 with respect to the background model, taking into account the first 14 bins.

{\bf \em  Production in the Sun followed by scattering.}---Now we consider a signal produced by scattering on electrons, which can recoil with 2-3 keV for light projectiles striking them with velocities $v/c>0.02$.
The model above is unsuitable for this signal as $v_\chi \lesssim 10^{-3} c$.
We thus consider an alternative possibility with light neutral fermions interacting with $e$ and $p$ via a dimension-6 operator:
\begin{align}
\mathcal{L} = \frac{1}{\Lambda^2} (\bar\psi p) (\bar\chi e)~.  
\end{align}
Constraints on this interaction come from $p$ decay (and potentially $^9$Be decay depending on the UV completion), as well as radiative H decay in Borexino~\cite{Agostini:2015oze,Alimonti:1998rc}, all of which are displayed in the bottom panel of Fig.~\ref{fig:tauH}.
For $m_\chi+m_\psi<m_{\rm H}$, this leads to the decay of hydrogen with
\begin{equation}
    \tau_{\rm H}\simeq 3\times 10^{27}~{\rm s}\sqrt{\frac{m_e}{Q}}\left(\frac{\Lambda}{100~\rm PeV}\right)^4,
\end{equation}
where here, $Q=m_{\rm H}-m_\chi-m_\psi\simeq m_p+m_e-m_\chi-m_\psi$. A large flux of $\chi$ and $\psi$ states comes from the decay of hydrogen in the Sun (or, more accurately, $e^-p$ capture in the stellar plasma). We assume that $\chi$ scatters on electrons, e.g. through a dimension-6 operator such as $\bar\chi\chi \bar e e$, and that the scattered electron gives a measurable signal. Post-production the $\chi$ and $\psi$ states move at speeds
\begin{align}
v_{\chi} = \sqrt{\frac{2 m_\psi Q}{m_\chi (m_\chi +m_\psi)}},
v_{\psi} = \sqrt{\frac{2 m_\chi Q}{m_\psi (m_\chi +m_\psi)}}~,
\end{align}
assuming $m_\chi$, $m_\psi \gg Q$.
Taking, for example, $m_\chi$=10~MeV and $Q = 100$~keV, we get $v_{\chi}/c = 0.14$, which, interestingly, provides a good fit to the excess at \acro{XENON}\osn{1}\acro{T} as seen in Ref.~\cite{Kannike:2020agf}.
The scattering rate is proportional to the incident flux, which is $O(10^3)$ higher for $ep$ capture in the Sun than in hydrogen decay in the oceans.
The solar flux of $\chi$ is given by~\cite{Berezhiani:2018udo}
\begin{align}
    \Phi^\odot_{\chi} = \frac{2.5 \times 10^{31}}{{\rm cm^{2}}\,\tau_{\rm H}} \sqrt{\frac{10^7 {\rm K}}{T}}
    \bigg( \frac{n_e}{2.5\times 10^{25} {\rm cm}^{-3}} \bigg)~,
    \label{eq:fluxsun}
\end{align}
where the electron number density $n_e$ and temperature $T$ have been normalized to their average values in the solar core, taken to have $0.2R_\odot$ radius.
The scattering rate of $\chi$ per ton of detector mass is 
\begin{align}
    \frac{dR}{dE_{\rm R}} = N_{\rm ton} \Phi^\odot_{\chi} \frac{d\sigma}{dE_{\rm R}}~,
\end{align}
where $N_{\rm ton}$ is the number of target atoms per ton of detector 
and the differential cross section is given by
\begin{align}
  \frac{d\sigma}{dE_{\rm R}} = \frac{\bar\sigma_e}{2\alpha^2 m_e^3 v^2_\chi}  \int^{q_+}_{q_-} dq q K(E_{\rm R},q)~, 
\end{align}
where $\bar\sigma_e$ is the cross section for scattering on unbound electrons at $q = \alpha m_e$, $K$ describes the probability to ionize the atom and the integration limits are
\begin{align}
    q_\pm = m_\chi v_\chi \pm \sqrt{m^2_\chi v^2_\chi - 2m_\chi E_{\rm R}}~.
\end{align}

Specifying now to the case of scattering in \acro{XENON}\osn{1}\acro{T}, we take the atomic ionization factor $K$, which for Xe is dominated by the $n=3$ shell, from Refs.~\cite{KfactorFlambaum,KfactorPospelov}, and assume a unit form factor, corresponding to a mediator with mass exceeding the momentum transfer $q = \mathcal{O}(10-100)$~keV. In Fig.~\ref{fig:signals} we display the event rate for a $\chi$ flux corresponding to $\tau_{\rm H} = 10^{27}$~s and $\bar\sigma_e=1.5\times 10^{-36}$~cm$^2$, seen to ameliorate the anomalous excess.
As before, we smear the signal rate with a 0.45~GeV Gaussian energy resolution and account for the detection efficiencies given in Ref.~\cite{X1TExcess}.
Using the first 14 bins we get $\chi^2$/d.o.f. = 0.58 with a 2-parameter fit, and $\Delta \chi^2 = 12.81$ with respect to the background model.

{\bf \em  Discussion.}---
Returning to our starting point with the maximum fluxes (\ref{fluxes}), we note that by virtue of very strong constraints on the proton lifetime, it is unlikely that proton decay could result in a sizeable 
rate for interactions/decays of daughter products in almost any realistic-sized detector. In contrast, it is is clear that dark matter decay 
could source relatively large interaction rates of its 
daughter products. In this paper, we  show that
hydrogen decay offers a unique window into new sources of exotic radioactivity that have largely gone unstudied. Large numbers of H atoms exist nearby in the Sun and on Earth, and could lead to appreciable signals in large volume detectors sensitive to energy depositions of 1-100~keV. The class of models where this could happen requires matching of the total mass of daughter particles to H mass to within $O(1\,{\rm MeV})$.

In this paper we have illustrated this point by explaining the recently reported \acro{XENON}\osn{1}\acro{T} electron recoil excess by $e^-p$ capture processes in the oceans and the Sun. 
In the first scenario, dark neutrons produced in hydrogen atom decays in the oceans in turn decay in-flight to a photon final state that deposits 2--3 keV energy in \acro{XENON}\osn{1}\acro{T}.
In the second scenario, long-lived dark sector states produced in $e^-p$ capture in the Sun travel at speeds $\gg 10^{-3}c$ and scatter on electrons in \acro{XENON}\osn{1}\acro{T}.
This scenario could also accommodate a signal from the decay of the dark sector particles produced in $ep$ capture, but we have not pursued this possibility.
The event rates in our model, displayed in Fig.~\ref{fig:signals}, provide a prediction for the event counts to expect at the future \acro{XENON}n\acro{T} experiment~\cite{xnt} with $\mathcal{O}(10)$ higher exposure than \acro{XENON}\osn{1}\acro{T}, as well as at \acro{PandaX}-\acro{II}~\cite{pandax}, \acro{lz}~\cite{lz}, and \acro{DarkSide}~\cite{darkside}.

We note that the solar flux in Eq.~\eqref{eq:fluxsun} scales as $T^{-1/2}$ due to Sommerfeld enhancement effects, whereas the flux of light boson states that contribute to stellar cooling are generally increasing functions of the temperature. In this setup, H decay in the sun contributes about one part in $10^7$ of the solar power radiated.
Moreover, constraints on light bosons from the cooling of red giants and horizontal branch stars are altogether avoided in this setup since the helium cores of these stars cannot decay due to nuclear binding energy. Further study of these and related models, particularly their effects on cosmology, would certainly be interesting.

{\it Acknowledgements.}--- D.\,M. and N.\,R. are supported in part by the Natural Sciences and Engineering Research Council of Canada (NSERC). TRIUMF receives funding through a contribution agreement with the National Research Council of Canada (NRC).
M.\,P. is supported in part by 
U.S. Department of Energy (Grant No. DE-SC0011842).

\bibliography{xe1t}

\end{document}